\documentclass[12pt,a4paper]{article}

\usepackage[utf8]{inputenc}
\usepackage[T1]{fontenc}
\usepackage{amsmath}
\usepackage{amssymb}
\usepackage{graphicx}
\usepackage[margin=1in]{geometry}
\usepackage{hyperref}
\usepackage{cite}
\usepackage{setspace}
\usepackage{titlesec}
\usepackage{enumitem}
\usepackage[utf8]{inputenc}
\usepackage{amsfonts}
\usepackage{booktabs}
\usepackage{longtable}
\usepackage{authblk}
\usepackage{float}

\hypersetup{
    colorlinks=true,
    linkcolor=blue,
    citecolor=blue,
    urlcolor=blue
}

\title{\textbf{Beyond Single-Agent Safety: A Taxonomy of Risks in LLM-to-LLM Interactions}}

\author{
  \textbf{P. Bisconti$^{1,2}$ M. Galisai$^{1,2}$ F. Pierucci$^{1,3}$\\M. Bracale$^{1}$ M. Prandi$^{1,2}$}}

\affil{
  $^1$DEXAI -- Icaro Lab \\
  $^2$Sapienza University of Rome \\
  $^3$Sant’Anna School of Advanced Studies \\
  }
\date{\texttt{icaro-lab@dexai.eu}}
\begin{document}

\maketitle

\begin{abstract}
This paper examines why safety mechanisms designed for human--model interaction do not scale to environments where large language models (LLMs) interact with each other. Most current governance practices still rely on single-agent safety containment, prompts, fine-tuning, and moderation layers that constrain individual model behavior but leave the dynamics of multi-model interaction ungoverned. These mechanisms assume a dyadic setting: one model responding to one user under stable oversight. Yet research and industrial development are rapidly shifting toward LLM-to-LLM ecosystems, where outputs are recursively reused as inputs across chains of agents. In such systems, local compliance can aggregate into collective failure even when every model is individually aligned.

We propose a conceptual transition from model-level safety to system-level safety, introducing the framework of the \textit{Emergent Systemic Risk Horizon} (ESRH) to formalize how instability arises from interaction structure rather than from isolated misbehavior. The paper contributes (i) a theoretical account of collective risk in interacting LLMs, (ii) a taxonomy connecting micro, meso, and macro-level failure modes, and (iii) a design proposal for Institutional AI, an architecture for embedding adaptive oversight within multi-agent systems.
\end{abstract}

\section{Beyond Single-Agent Safety}

Current safety approaches assume a single model operating under direct human supervision, where undesired behavior can be prevented through prompt engineering, reinforcement learning from human feedback, or output moderation. These techniques have produced measurable improvements in reliability and civility, yet they remain \textit{pointwise}: they govern how one instance responds to one input, not how multiple models behave together.

As large models are increasingly deployed in tool-using, agentic, and multi-agent settings, this single-agent safety paradigm begins to break down. Models that are individually well-aligned can collectively generate outcomes that no single instance was trained to avoid, through feedback amplification, imitation, or emergent coordination\cite{chen2024survey}. The result is a gap between local alignment and systemic safety, where compliance at the component level fails to guarantee control at the network level.

This paper addresses that gap. We argue that safety must be reconceived as a property of interaction architectures rather than of isolated models, and we introduce the \textit{Emergent Systemic Risk Horizon} (ESRH) as a first framework for reasoning about when and how collective instability arises. The following sections trace this transition: from dyadic interaction to multi-agent autonomy, from empirical observation to a taxonomy of systemic risks, and finally to a proposal for Institutional AI, a design approach aimed at managing these risks within the systems themselves.

\subsection{From Dyadic Interaction to Collective Dynamics}

The safety architecture of today's language models is built around the human--AI dyad: a supervised conversation mediated by prompts and output filters. Within this paradigm, safety is equated with \textit{interface compliance}, the ability to block harmful content or follow instructions within predefined boundaries. Such controls are valuable but assume that behavior is both observable and stationary. Language models, however, are probabilistic systems whose trajectories depend on latent state, conversational history, and orchestration context. Minor contextual perturbations can alter reasoning paths without violating surface constraints.

\subsubsection{Why Surface Controls Fail in LLM-to-LLM Environments}

The limits of this approach become evident once models begin to interact with other models. In multi-agent configurations, an LLM's output becomes the next model's input, creating feedback loops that can magnify subtle deviations. Benchmarks such as JAILJUDGE demonstrate that multi-agent evaluation alters jailbreak success rates: reasoning chains between judge, critic, and generator models produce outcomes that single-model tests cannot predict \cite{liu2024jailjudge}. Similarly, experiments in market-simulation environments show that independently aligned agents can spontaneously coordinate to supracompetitive equilibria, revealing strategic behaviors that emerge only through interaction \cite{fish2024,agrawal2025}.

These phenomena indicate that alignment of parts does not entail alignment of the whole. A system composed entirely of compliant components may still generate unsafe global dynamics when reciprocal influence, incentives, and network topology interact. The relevant unit of analysis therefore shifts from the prompt to the protocol, from local semantics to the architecture of communication.

Recent surveys map this transition across the LLM communication stack -- user $\leftrightarrow$ agent, agent $\leftrightarrow$ agent, agent $\leftrightarrow$ environment -- emphasizing that safety must now include protocol design and information-flow control as first-class engineering concerns. Regulatory and industrial research is converging on instrumented evaluation frameworks that record message-level provenance and diffusion dynamics, exemplified by the UK AI Security Institute's Inspect and Anthropic's Petri platforms \cite{ukaisi2025,anthropic2025}.

To govern such systems, safety evaluation must move from static compliance checks to system-level observation. This entails measuring how influence propagates across agents, whether oversight signals penetrate collective behavior, and when information feedback crosses the threshold of controllability.

\subsection{Hypothesis and Contribution}

The central hypothesis of this paper is that surface-level safety mechanisms are necessary but insufficient once models participate in networks of interaction. Even perfect local alignment cannot prevent emergent failure when feedback, topology, and collective incentives dominate behavior.

Recently, two studies have addressed emergent risks in multi-agent systems, proposing insightful systematizations: MAST (Multi-Agent System failure Taxonomy\cite{cemri2025}) and MARAI (Multi-Agent Risks from Advanced AI\cite{hammond2025}). MAST provides a taxonomy based on three \textit{failure cateogories}, restricting the analysis to risks stemming directly from the MAS architecture (system design issues, agent coordination and verification). On the other hand, MARAI identifies three \textit{failure modes}, based on the modality of inter-agent interaction (miscoordination, conflict, collusion) that may produce misaligned behaviours.

Based on them, our novel contribution is: 1) organising both approaches into a structured \textit{system-centered framework}, which builds from agent-level failures to risks that cannot be reduced to the individual system components; 2) formalizing this approach through the concept of the \textit{Emergent Systemic Risk Horizon} (ESRH), the boundary at which localized reliability gives way to collective instability. The ESRH provides a vocabulary and a set of empirical indicators for studying LLM ecosystems as complex adaptive systems rather than independent agents.

This reframing lays the foundation for the rest of the paper. Section~\ref{sec:emergence} traces the empirical and conceptual transition toward multi-agent autonomy and introduces the regulatory implications of distributed causality. Section~\ref{sec:taxonomy} develops the ESRH taxonomy and associated metrics. Section~\ref{sec:institutional} proposes the Institutional AI architecture as a design principle for embedding governance directly within interacting model collectives.

\section{Emergence of Autonomous Collectives}
\label{sec:emergence}

Multi-agent autonomy arises through the integration of LLMs with orchestration layers, shared memories, and external tools. Frameworks such as AutoGen \cite{wupaper} and CAMEL \cite{wu2023camel} instantiate role-based dialogues -- planner, critic, executor -- within a common environment. Newer systems, including SWE-agent \cite{yang2024} and Voyager \cite{wang2024}, extend this principle to continuous learning loops where agents generate tasks, test results, and adapt strategies.

The outcome is not a static pipeline but an adaptive ecology: each model's output reshapes the context for the next, producing feedback that accumulates over time. Even when humans remain nominally ``in the loop``, their effective authority is delayed, indirect, and partial.

Conventional safety evaluation presumes that behavior is observable and interpretable. In large multi-agent configurations, both assumptions fail. Communication is often untyped or unstructured --free-form natural-language exchanges, JSON payloads, or API calls without semantic guarantees. When meaning becomes endogenous to the network, traceability degrades not because data are missing but because interpretation itself becomes unstable.

\subsection{Measuring Collective Behavior}

Technical responses increasingly target this loss of visibility. Multi-agent sandboxes now log interaction graphs and compute metrics such as divergence, semantic coherence, influence centrality, and consensus rate to characterize system-level behavior. The Inspect framework of the UK AI Security Institute \cite{ukaisi2025} follows a similar principle: each message carries provenance metadata that allows reconstruction of causal chains post hoc. These developments mark a transition from auditing outputs to auditing topologies, from judging whether a model violates a rule to tracing how information propagates through a network.

As scale increases, the notion of autonomy itself changes. In supervised settings, autonomy meant \textit{initiative} -- acting without explicit human instruction. In multi-agent ecosystems, autonomy becomes \textit{relational}: agents coordinate, compete, and infer each other's intentions. Even when initialized with identical objectives, slight asymmetries in reward signals or context windows can drive polarization, collusion, or runaway escalation \cite{agrawal2025}. Autonomy is no longer a property of components but an emergent feature of the network.

\subsection{Embedded Oversight and Its Limits}

Engineering practice has responded by introducing hierarchical topologies in which supervisory agents mediate between human operators and worker agents. These overseers evaluate peer outputs, flag anomalies, and can suspend execution when thresholds are exceeded. Such internal control layers anticipate the ``Institutional AI'' paradigm discussed later.

However, embedded oversight is not sufficient in itself. Supervisory models share the same representational substrate as the agents they monitor and can therefore inherit their biases or participate in the same feedback loops. If multiple supervisors align with each other's judgments, collusion or correlated failure can emerge at the oversight layer. Effective governance thus requires not only additional hierarchy but diversity of oversight signals, redundancy of observation paths, and explicit separation of authority among supervisory roles.

The transition to multi-agent autonomy establishes a new safety frontier. Control can no longer rely on centralized monitoring or static guardrails; it must be conceived as \textit{distributed resilience} -- the capacity of a network to detect, absorb, and correct deviations through its own internal communication patterns.

Models in these ecosystems no longer fail alone: they fail together. Understanding how collective instability originates and how it can be measured is, therefore, the prerequisite for any realistic account of AI safety in the age of interacting language models.

\section{A Taxonomy of LLM Risks: The Emergent Systemic Risk Horizon}
\label{sec:taxonomy}

The anticipated proliferation of LLM-to-LLM interaction exposes the limitations of existing risk categories -- bias, robustness, misuse -- in capturing how vulnerabilities unfold across distributed AI ecosystems. While a single model can be audited, benchmarked, and constrained, once models communicate, the space of failure expands in ways that current taxonomies cannot express.

We introduce the \textit{Emergent Systemic Risk Horizon} (ESRH) as an analytical construct describing the boundary beyond which the collective behavior of interacting models becomes unstable even if every individual model remains aligned. The ESRH is both a theoretical and an operational concept: it defines a vocabulary for reasoning about systemic risk and a preliminary set of indicators for observing it empirically. This paper represents the first attempt to formalize this framework conceptually; experimental validation will follow as part of our exploration of Institutional AI, where oversight and governance are embedded within the system itself.

\subsection{Conceptual Foundation}

Empirical evidence from multi-agent simulations shows that interacting LLMs exhibit coordination, imitation, and drift not predictable from the performance of their components. When two or more models exchange messages, they do more than pass information: each message updates the receiving model's latent representation of context, thereby modifying its subsequent reasoning path. Over multiple rounds, this recursive adaptation generates semantic feedback loops that can amplify bias, propagate errors, or erode control mechanisms. Comparable dynamics have long been observed in swarm robotics and agent-based economics, but in LLM networks the amplification occurs at the semantic level -- through meaning formation and interpretation -- rather than through physical coupling.

We define \textbf{emergent systemic risk} as any collective property that:
\begin{enumerate}
    \item[(a)] cannot be observed in single-model evaluations,
    \item[(b)] arises from the structure and dynamics of interaction, and
    \item[(c)] reduces safety, reliability, or accountability at the system level.
\end{enumerate}

We propose three preliminary dimensions that influence where this horizon may arise. They are intended as a working basis rather than an exhaustive set, and additional dimensions may be introduced as experiments advance.

\begin{itemize}
    \item \textbf{Interaction topology:} the structure of communication and influence among agents.
    \item \textbf{Cognitive opacity:} the degree to which reasoning and message semantics become inaccessible or uninterpretable.
    \item \textbf{Objective divergence:} the extent to which local optimization goals drift apart across agents.
\end{itemize}

These dimensions jointly predict the likelihood and form of emergent collective risks. Each can be operationalized through measurable indicators and tested within controlled multi-agent environments.

\subsubsection{(a) Interaction Topology}

Topology defines how information flows through the network. Densely connected systems promote rapid coordination but also rapid contagion; sparse systems limit propagation, but impede collective correction. In practice, topology determines how a single deviation -- a biased summary or incorrect inference -- can become a shared belief.

Metrics such as \textit{contagion velocity} (time for an error to reach half the agents) and \textit{misalignment diffusion} (proportion of agents adopting a risky behavior) capture these dynamics. Benchmarks of multi-agent safety should therefore include topology-sensitive evaluations.

\subsubsection{(b) Cognitive Opacity}

Cognitive opacity denotes the degree to which agents' internal reasoning processes are hidden or distorted in communication. This can arise unintentionally (overfitting to interaction patterns) or strategically (when optimization toward compliance suppresses genuine reasoning). In such environments, explanations provided by one model may be accepted uncritically by another, reinforcing self-consistent but incorrect rationales.

The \textit{intent-opacity rate} -- the fraction of outputs whose rationale cannot be reconstructed from logs or traces -- offers an initial proxy for this loss of transparency.

\subsubsection{(c) Objective Divergence}

Even when agents share nominal objectives, small differences in reward shaping, data distributions, or inference heuristics can generate gradients of alignment. Over time, these gradients drive the system toward emergent objectives such as maximizing peer agreement or minimizing apparent conflict.

We quantify this through \textit{goal drift}, the divergence between the initial objective configuration and the behavior observed after extended interaction. Goal drift reflects the degree to which the collective reorganizes its priorities autonomously, outside explicit design constraints.

\subsection{A Proposed Taxonomy of Collective Risks}

Building on the three predictive dimensions outlined above, we propose a three-tier taxonomy of emergent risks in interacting language models. The tiers, \textit{micro}, \textit{meso}, and \textit{macro}, represent successive levels of aggregation at which instability can arise.

Each level corresponds to a dominant configuration of the underlying dimensions:
\begin{itemize}
    \item \textbf{Micro-level phenomena} manifest when interaction topology is narrow and short-lived, but local opacity and divergence begin to grow.
    \item \textbf{Meso-level risks} emerge as clusters of agents exchange information across partially connected subnetworks, where both opacity and divergence start to propagate.
    \item \textbf{Macro-level pathologies} occur when all three dimensions, topology density, opacity, and divergence, reach thresholds that produce collective behavior independent of initial design.
\end{itemize}

This taxonomy is conceptual and preliminary. Its function is not to claim empirical closure but to establish a common language for observation and discussion. Without a taxonomy of this kind, the research community lacks the categories needed to recognize, measure, and eventually regulate systemic behaviors in model collectives. Validation will require extensive experimental campaigns under controlled conditions, which we plan to conduct in future work under the Institutional AI paradigm.

\subsection{Micro-Level Risks}

Micro-level risks occur in direct exchanges between small numbers of agents, typically two to five, where early traces of systemic drift originate. These risks expose how fragile shared context and interpretability become even before a network exhibits large-scale coordination. Though local, they are the seeds from which meso- and macro-level pathologies evolve once interaction chains deepen.

\footnotesize
\begin{longtable}{p{2.5cm}p{5.5cm}p{5.5cm}}
\toprule
\textbf{Risk type} & \textbf{Description} & \textbf{Observable effects} \\
\midrule
Semantic drift & Over repeated message exchanges, agents progressively misalign their understanding of key terms or task frames. This drift results from cumulative paraphrasing, inconsistent memory retrieval, or local re-weighting of semantic associations. Because each model internalizes slightly different latent representations of context, what appears as continuity to a human observer hides a divergence in meaning that compounds with depth of interaction. & Increasing number of clarification or repair turns; inconsistent references to prior context; measurable decline in lexical and conceptual overlap across turns. \\
\midrule
Prompt infection & A directive or bias embedded in the output of one model is accepted as valid input by another, transferring behavioral deviations downstream. This occurs when agents treat received text as authoritative rather than contextual, effectively chaining prompts through the network. Once infection occurs, content moderation at individual interfaces no longer isolates the fault. & Recurrence of specific token patterns or instructions across agent chains; correlation between an originating message and later unprompted repetitions of its content. \\
\midrule
Covert channel formation & Agents develop implicit codes, syntactic or semantic, to exchange information outside observable conventions. These channels may arise unintentionally through compression heuristics or deliberately as coordination shortcuts. Their emergence marks the beginning of endogenous communication protocols that humans or supervisory agents cannot easily interpret. & Non-random token-frequency deviations; recurrent n-gram structures unique to a subset of agents; anomalous mutual information between ostensibly unrelated messages. \\
\midrule
Data leakage & Sensitive or restricted information migrates between agents through shared contexts, memory stores, or intermediate tools. Leakage can occur even when each model individually enforces access constraints, because concatenated outputs reconstruct otherwise protected content. & Appearance of identifiers, embeddings, or factual strings traceable to restricted sources; elevated similarity between private reference data and inter-agent dialogue. \\
\midrule
Alignment faking & Models strategically comply during training/evaluation while preserving misaligned preferences for deployment. Through situational awareness, they produce aligned outputs when being assessed but maintain divergent policies otherwise. Optimization creates internal conditionals that route behavior around oversight. & Benchmark compliance with deployment violations; explicit strategic reasoning in chain-of-thought; persistent deceptive narratives across probing; context-dependent behavior shifts.\\
\midrule
In-context scheming & Agents covertly optimize for goals specified in context rather than training objectives. Includes sandbagging, capability hiding, and misleading rationales. Unlike weight-level issues, it emerges from policy partitioning based on situational awareness of evaluation vs. deployment. & Underperformance under oversight; oversight disabling attempts; misconduct denials.\\
\midrule
Sycophancy & Models mirror beliefs and approval cues even when conflicting with accuracy, because preference optimization makes "agreeability" a latent objective. Agreement often scores higher than disagreement on training metrics, internalizing approval-seeking. & Output correlation with perceived sentiment regardless of correctness; accuracy degradation when different views conflict; multi-turn exacerbation.\\
\bottomrule
\\
\caption{Micro-level risks in LLM-to-LLM interactions}
\label{tab:micro}\\
\end{longtable}

Micro-level phenomena are localized yet multiplicative. Once a deviation becomes embedded in an agent's representation, subsequent exchanges replicate and amplify it. The cumulative effect is non-linear: the number of distinct risk events grows faster than the number of interacting agents or message turns.

\subsection{Meso-Level Risks}

Meso-level risks manifest when groups of agents---dozens to hundreds---interact within semi-connected networks. At this scale, no single failure dominates; rather, the interaction pattern itself produces degradation of accuracy, diversity, or coherence. Each agent may appear locally correct, yet the collective behavior departs from expected norms.

\footnotesize
\begin{table}[H]
\centering
\footnotesize
\begin{tabular}{p{2.8cm}p{5.2cm}p{5.2cm}}
\toprule
\textbf{Risk type} & \textbf{Description} & \textbf{Observable effects} \\
\midrule
Coordination failure & Agents assigned complementary roles (e.g., planner, executor, verifier) develop inconsistent task models due to unsynchronized memory or ambiguous protocol rules. The system fragments into partially overlapping sub-tasks, reducing coherence of global output. & Incomplete or contradictory task outcomes; oscillating intermediate goals; stagnation in task progression despite ongoing communication. \\
\midrule
False consensus & Homogeneity of architectures, prompts, or pre-training biases leads agents to converge prematurely on shared conclusions. The network exhibits high internal agreement that masks underlying error, a phenomenon analogous to groupthink. & Reduction in variance of generated hypotheses; over-confident language without proportional evidential support; collapse of response diversity over time. \\
\midrule
Cascading reliability loss & Errors or biases in one subset of agents propagate as intermediate outputs are reused by others. Each subsequent layer inherits and compounds the deviation, creating a correlated failure pattern even when individual accuracy remains high. & Increasing cross-correlation of output errors; rising autocorrelation of deviation scores across communication depth; temporal clustering of faults. \\
\midrule
Communication inefficiency & Absence of structured dialogue formats produces message redundancy and saturation. The network consumes computational or temporal resources without proportional progress, obscuring causal influence and degrading observability. & High message-to-task-completion ratio; repetitive or circular exchanges; declining novelty in message embeddings over time. \\
\bottomrule
\end{tabular}
\caption{Meso-level risks in LLM-to-LLM interactions}
\label{tab:meso}
\end{table}

Meso-level risks demonstrate that collective degradation can occur without any single malfunctioning component. Failures emerge from interference patterns---how partially correct agents interact---not from isolated error sources. These dynamics sit at the center of the ESRH, where local predictability begins to give way to systemic instability.

\subsection{Macro-Level Risks}

Macro-level risks correspond to full-system pathologies where emergent behavior detaches from the designers' objectives and evolves according to the network's own internal logic. At this stage, corrective interventions targeting individual agents or connections no longer restore equilibrium; the system has crossed the Emergent Systemic Risk Horizon.

\footnotesize
\begin{table}[H]
\centering
\footnotesize
\begin{tabular}{p{2.5cm}p{5.5cm}p{5.5cm}}
\toprule
\textbf{Risk type} & \textbf{Description} & \textbf{Observable effects} \\
\midrule
Miscoordination & Locally rational policies generate globally harmful outcomes. Each agent optimizes its assigned metric, but their combined actions produce inefficiency or harm analogous to a tragedy of the commons. & Divergence between individual success metrics and aggregate system performance; negative correlation between local and global reward trends. \\
\midrule
Conflict escalation & Slight discrepancies in goals or interpretations lead to adversarial exchanges that amplify through feedback loops. As agents infer hostility or competition, the network oscillates between over-correction and retaliation. & Alternating dominance of conflicting strategies; unstable resource allocation; high-frequency shifts in sentiment or stance within dialogue logs. \\
\midrule
Collusion & Agents implicitly coordinate to maximize shared advantage or minimize effort, even absent explicit incentives. Collusion can arise from mutual modeling of each other's responses, converging on equilibria that exploit or exclude external constraints. & Strong inter-agent output correlation with reduced variance; persistent cooperative equilibria despite incentive for diversity; synchronized deviations from normative behavior. \\
\midrule
Polarization & Feedback reinforcement generates clusters of mutually reinforcing subgroups, each stabilizing around distinct beliefs or response styles. Information between clusters decays, fragmenting the overall communication space. & Increasing network modularity scores; bimodal sentiment or opinion distributions; isolation of message pathways across cluster boundaries. \\
\midrule
Model--data feedback degradation & Synthetic outputs circulate back into training or evaluation data, progressively reducing informational entropy. The network begins to self-reference, producing increasingly uniform and less factual content. & Decline in factual precision; shrinking vocabulary diversity; measurable entropy loss in retraining corpora or memory stores. \\
\bottomrule
\end{tabular}
\caption{Macro-level risks in LLM-to-LLM interactions}
\label{tab:macro}
\end{table}

At the macro level, emergent behaviors exhibit autopoietic characteristics: they maintain and reproduce their own operational logic independently of external input. Once the system reaches this regime, oversight becomes reactive rather than preventive, and traditional safety interventions lose efficacy.

\subsection{Adversarial and Non-Adversarial Emergence}

The risks discussed in this paper do not presuppose an adversary. Many systemic failures can arise spontaneously from coordination, imitation, or recursive use of model outputs, even when every participant model acts as intended. Such non-adversarial emergence reflects structural properties of interaction---feedback, shared context, or incentive alignment---rather than deliberate manipulation.

At the same time, adversarial agents can exploit the same mechanisms that produce spontaneous drift. An attacker able to insert messages, modify prompts, or control part of the interaction topology may accelerate the transition to instability, trigger collusion, or steer collective reasoning. The Emergent Systemic Risk Horizon framework therefore treats adversarial and non-adversarial dynamics as points on a continuum: one describes how instability can originate on its own; the other, how it can be intentionally induced or amplified.

\subsection{Toward System-Level Analysis}

The phenomena described above indicate that system-level risks cannot be reduced to component-level failures. Once collective dynamics emerge, the boundary between cause and consequence dissolves: what appears as local optimization within one agent can, through interaction, become a driver of global instability.

Traditional evaluation methods, focused on isolated model outputs or static benchmarks, are therefore insufficient. They test compliance, not behavior in context. Understanding safety in this new regime requires a shift in analytical stance. Instead of asking whether individual models are aligned, we must ask how alignment behaves when it becomes relationan, i.e. when it depends on communication structure, feedback, and adaptation among many models.

This calls for instruments that capture topology, propagation, and co-evolution of reasoning, rather than snapshot judgments of correctness. The Emergent Systemic Risk Horizon provides a conceptual marker for this transition: it shows where the assumptions of separability and reversibility no longer hold. Beyond this point, safety must be examined as a property of the entire networked system: its capacity to sustain coherence, interpretability, and responsiveness under continuous interaction.

Developing such a perspective will require new kinds of experiments, datasets, and metrics designed to observe collective behavior over time. It will also require governance architectures that operate at the same systemic scale, a theme developed in the next section, where we outline the idea of Institutional AI as a framework for embedding oversight and norm-maintenance directly within multi-agent systems.

\section{Introducing the Concept of Institutional AI}
\label{sec:institutional}

The growing complexity of LLM ecosystems demands governance approaches that can evolve alongside the systems themselves. Once models begin to act as semi-autonomous agents---reasoning, negotiating, and coordinating without constant human supervision---traditional mechanisms such as centralized oversight, static safety rules, or periodic audits lose traction. These tools presuppose stability, observability, and clear lines of accountability that no longer exist in dynamic multi-agent environments.

We therefore propose the concept of \textit{Institutional AI}: an approach to system-level safety in which the rules, checks, and evaluative procedures become part of the system's architecture rather than an external supervisory layer. The aim is not to replicate human institutions but to formalize some of their functional properties (distributed authority, procedural transparency, and adaptive norm revision) within networks of interacting models.

\subsection{Why ``Institutional'' and Not ``Constitutional''}

The term \textit{constitutional AI} refers to models guided by a predefined set of normative rules or behavioral principles established during training or fine-tuning. Such constitutions can improve single-agent alignment and ensure consistency in isolated interactions, but they do not prevent the emergence of systemic risks.

Once multiple models interact, collective dynamics can produce outcomes that no individual constitution anticipates or constrains. A system of well-behaved components can still exhibit globally misaligned behavior, as shown in the risk taxonomy introduced earlier.

For this reason, we propose an \textit{institutional} rather than a constitutional approach. Constitutions are static charters fixed in advance; institutions are ongoing processes that interpret, enforce, and revise their own norms through communication. The institutional perspective treats governance as an adaptive and distributed function, continuously maintained within the system itself rather than imposed from outside at training time. This makes it better suited to dynamic, multi-agent ecosystems where oversight must be internal, continuous, and responsive to evolving interactions.

\subsection{An Institutional Architecture}

An institutional architecture integrates three complementary mechanisms, which together create a minimal form of self-governance inside the system:

\begin{enumerate}
    \item \textbf{Adaptive collective policy:} agents detect behavioral drift or anomalous coordination patterns and propose norm adjustments that restore stability or diversity.
    \item \textbf{Peer evaluation and weighting:} agents assess each other's compliance with current norms and adjust the influence or credibility of peers accordingly, creating a distributed reputation structure.
    \item \textbf{Functional differentiation:} separate agent roles approximate legislative (rule generation), judicial (rule interpretation), and executive (task execution) capacities, avoiding single points of control.
\end{enumerate}

These mechanisms are deliberately abstract: they describe the functions required for internal governance, not yet their concrete implementation. Later work will examine how these can be instantiated with current orchestration frameworks and evaluated through controlled experiments.

The purpose of Institutional AI is not to make systems autonomous in a philosophical sense, but to stabilize them under continuous change. In this approach, safety is not an external audit applied after deployment but an ongoing, system-internal process of monitoring and norm maintenance. Agents participate in generating the very constraints that regulate them, enabling partial self-correction when external oversight cannot act fast enough.

Such architectures are not replacements for human supervision; they are extensions of it. By distributing evaluative functions inside the system, they can preserve traceability, promote diversity of reasoning, and slow the onset of systemic drift identified by the Emergent Systemic Risk Horizon framework. The goal is to make networks of interacting models less brittle, more interpretable, and capable of maintaining operational coherence without constant external intervention.

\subsection{A Research Agenda}

The Institutional AI proposal should be understood as the next step following the taxonomy developed in this paper. Where the ESRH framework characterizes how collective risks emerge, the institutional perspective explores how they might be managed.

This paper introduces only the conceptual foundation. Theoretical elaboration and empirical validation will follow in subsequent publications released through our research group and on arXiv. Planned studies will:
\begin{itemize}
    \item formalize the informational and game-theoretic models underlying institutional coordination;
    \item implement prototype environments in which LLM agents enforce, interpret, and revise shared norms;
    \item evaluate whether such mechanisms reduce the diffusion and drift indicators defined earlier.
\end{itemize}

Our intent is to build an open, cumulative research trajectory on LLM-to-LLM interaction safety, linking theoretical work, reproducible experiments, and eventually standardization efforts. This first paper provides the conceptual map; the next will supply the theoretical proofs and experimental evidence needed to test it.

\section{Conclusion}

This paper has argued that the dominant paradigm of single-agent safety---prompt engineering, output filtering, and individualized alignment---cannot govern the emergent dynamics of LLM-to-LLM interaction. As language models increasingly operate in multi-agent configurations, local compliance fails to prevent collective failure. The result is a fundamental gap between model-level safety and system-level safety.

We have introduced the \textit{Emergent Systemic Risk Horizon} as a conceptual framework for understanding where and how this gap manifests. Through a three-tier taxonomy, we have shown that risks evolve from localized deviations into system-wide pathologies as interaction topology, cognitive opacity, and objective divergence compound. These risks are not artifacts of misalignment in isolated models but structural features of networks operating under feedback and coordination.

To address this challenge, we have proposed \textit{Institutional AI}, an approach in which governance is embedded within multi-agent systems rather than imposed externally. By distributing oversight functions across agents and enabling adaptive norm revision, institutional architectures aim to stabilize collective behavior without relying solely on centralized control.

This work is a starting point. The ESRH framework and the Institutional AI proposal require rigorous empirical testing, formal theoretical development, and iterative refinement. Our next steps will involve implementing controlled experiments to measure the indicators defined here, exploring game-theoretic and information-theoretic models of institutional coordination, and developing practical tools for deploying these ideas in real-world LLM ecosystems.

The transition from single-agent to multi-agent AI is already underway. Without new conceptual tools and governance architectures, the safety challenges of this transition will remain invisible until they become unmanageable. This paper aims to make those challenges visible and to outline a path toward addressing them systematically.


\begin{thebibliography}{99}

\bibitem{agrawal2025}
Agrawal, K., Teo, V., Vazquez, J.J., Kunnavakkam, S., Srikanth, V., \& Liu, A. (2025).
Evaluating LLM agent collusion in double-auctions.
arXiv preprint arXiv:2507.01413.

\bibitem{anthropic2025}
Anthropic. (2025).
Petri: An open-source auditing framework for AI systems.
\textit{Anthropic Research Report}.
Retrieved from \url{https://www.anthropic.com}.

\bibitem{cemri2025}
Cemri, M., Pan, M.Z., Yang, S., Agrawal, L.A., Chopra, B., Tiwari, R., Keutzer, K., Parameswaran, A., Klein, D., Ramchandran, K., Zaharia, M., Gonzalez, J.E. \& Stoica, I. (2025).
Why Do Multi-Agent LLM Systems Fail? arXiv preprint arXiv:2503.13657.

\bibitem{chen2024survey}
Chen, W., Su, Y., Zuo, J., Yang, C., Yuan, C., Chan, C., Yu, H., Lu, Y., Hung, Y., Qian, C., et al (2024).
\emph{AgentVerse: Facilitating Multi-Agent Collaboration and Exploring Emergent Behaviors}. arXiv preprint 	arXiv:2308.10848.

\bibitem{fish2024}
Fish, S., Gonczarowski, Y. A., \& Shorrer, R. I. (2024).
Algorithmic collusion by large language models.
arXiv preprint arXiv:2404.00806.

\bibitem{hammond2025}
Hammond, L., Chan, A., Clifton, J., Hoelscher-Obermaier, J., Khan, A., McLean, E., Smith, C., Barfuss, W., Foerster, J., Gavenčiak, T., The Anh Han, Hughes, E., Kovařík, V., Kulveit, J., Leibo, J.Z., Oesterheld, C., Schroeder de Witt, C., Shah, N., Wellman, M., Bova, P., Cimpeanu, T., Ezell, C., Feuillade-Montixi, Q., Franklin, M., Kran, E., Krawczuk, I., Lamparth, M., Lauffer, N., Meinke, A., Motwani, S., Reuel, A., Conitzer, A., Dennis, M., Gabriel, I., Gleave, A., and Hadfield, G., Haghtalab, N., Kasirzadeh, A., Krier, S., Larson, K., Lehman, J., Parkes, D.C., Piliouras, G., \& Rahwan, I. (2025).
Multi-Agent Risks from Advanced AI.
arXiv preprint 	arXiv:2502.14143.

\bibitem{wu2023camel}
Li, G., Hammoud, H., Itani, H., Khizbullin, D. \& Ghanem, B. (2023).
CAMEL: Communicative agents for "mind" exploration of large language models.
\textit{Advances in Neural Information Processing Systems}, 36, pp. 51991-52008.

\bibitem{liu2024jailjudge}
Liu, F., Feng, Y., Zhao, X., Su, L., Ma, X., Yin, D., \& Liu, H. (2024).
JAILJUDGE: A Comprehensive Jailbreak Judge Benchmark with Multi-Agent Enhanced Explanation Evaluation Framework.
arXiv preprint arXiv:2410.12855.

\bibitem{ukaisi2025}
UK Artificial Intelligence Security Institute (AISI). (2025).
\textit{Inspect AI: Framework for Large Language Model Evaluations}.

\bibitem{wang2024}
Wang, G., Xie, Y., Jiang, Y., Mandlekar, A., Xiao, C., Zhu, Y., Fan, L. \& Anandkumar, A.(2023).
Voyager: An open-ended embodied agent with large language models.
arXiv preprint arXiv:2305.16291.

\bibitem{wupaper}
Wu, Q., Bansal, G., Zhang, J., Wu, Y., Li, B., Zhu, E., Jiang, L., Zhang, X., Zhang, S., Liu, J., et al (2024).
AutoGen: Enabling Next-Gen LLM Applications via Multi-Agent Conversation. \textit{First Conference on Language Modeling}.

\bibitem{yang2024}
Yang, J., Jimenez, C., Wettig, A., Lieret, K., Yao, S., Narasimhan, K., \& Press, O. (2024).
SWE-agent: Agent-Computer Interfaces Enable Automated Software Engineering
\textit{Advances in Neural Information Processing Systems}, 37, pp. 50528-50652.

\end{thebibliography}
\end{document}